\documentclass[12pt]{article}
\usepackage{amsxtra,amssymb,amsthm,amsmath,latexsym}

\theoremstyle{plain}

\newtheorem{theorem}{Theorem}[section]
\newtheorem{proposition}{Proposition}[section]

\newtheorem{lemma}{Lemma}[section]

\newtheorem{remark}{Remark}[section]

\def\R{{\mathbb R}}

\def\C{{\mathbb C}}
\def\oH{\buildrel\circ\over H}
\def\oH1{\buildrel\circ\over H\kern-.02in{}^1}
\def\dotf{\buildrel\dot\over f}

\def\l{\ell}
\def\dotf{\dot{f}}
\def\tildeH{\widetilde H}
\def\tildev{\widetilde v}
\def\be{\begin{equation}}
\def\ee{\end{equation}}
\def\ind{\hbox{ind}}

\def\bysame{\rule{.5in}{.005in}}
\def\Im{\hbox{\,Im\,}}

\begin{document}


\title{  Krein's method in inverse scattering                
  \thanks{The work on this paper was started while the author
visited
     BGU in Beer-Sheva. The author thanks BGU for hospitality
     and Professor D. Alpay for useful discussions.}
   \thanks{key words:   Inverse scattering, Krein's method}
   \thanks{Math subject classification:   34B25, 34R30,
PACS 02.30.Hq 03.40.Kf,  03.65.Nk}
}

\author{
A.G. Ramm\\
 Mathematics Department, Kansas State University, \\
 Manhattan, KS 66506-2602, USA\\
ramm@math.ksu.edu\\
}

\date{}

\maketitle\thispagestyle{empty}

\begin{abstract}
A detailed discussion of the Krein's results ( applicable for
solving the inverse scattering
problem) is given with complete proofs.

It is shown that the $S$-function $S(k)$ 
used in Krein's work
is the $S$-matrix used
in physics. 

The basic new results of the paper include
the detailed description
and analysis of an inversion algorithm based on Krein's
results  and a proof of its consistency, that is the proof
that the reconstructed potential generates the same 
scattering data from which it was reconstructed.

Numerical advantages of using Krein's method are discussed.
\end{abstract}


\section{Introduction}

The inverse scattering problem on half-axis consists of finding
$$q(x)\in L_{1,1}=\left\{q:q=\overline q,
  \quad \int^\infty_0x |q(x)|dx<\infty\right\}$$
from the knowledge of the scattering data
\begin{equation}
  {\cal S}:=\{S(k),k_j,s_j,1\leq j\leq J\}.
  \tag{1.1} \end{equation}
Here
$$ S(k):=\frac{f(-k)}{f(k)} $$
is the $S$-matrix, $f(k)$ is the Jost function
\begin{equation}
  f(k):=f(0,k),
  \tag{1.2} \end{equation}
$f(x,k)$ is the solution to the equation
\begin{equation}
 \l f:=f^{\prime\prime}+k^2f-q(x)f=0
 \tag{1.3} \end{equation}
which is uniquely defined by the condition
\begin{equation}
  f(x,k)=e^{ikx}+o(1),\quad x\to+\infty,
  \tag{1.4} \end{equation}
$k_j>0$ are the (only) zeros of $f(x)$ in the region
$\C_+:=\{k:\Im k>0\}$, $-k_j^2$ are the negative eigenvalues
of the Dirichlet operator
$-\frac{d^2}{dx^2}+q(x)$ on $(0,\infty)$, $s_j>0$ are the norming
constants:
\begin{equation}
  s_j=-\frac{2ik_j}{\dotf(ik_j)f^\prime(0,ik_j)},
  \tag{1.5}\end{equation}
and $J\geq 0$ is the number of the negative eigenvalues.

For simplicity {\it we assume that there are no bound states.}
This assumption is removed in section 4.

This paper is a commentary to Krein's paper \cite{K1}.
It contains not only a detailed proof of the results announced in
\cite{K1} but also a proof of the new results not mentioned in
\cite{K1}.
In particular, it contains an analysis of the invertibility of the
steps in the inversion procedure based on Krein's 
results, and a proof of
the consistency of this procedure, that is, a proof of the fact
that the reconstructed potential generates the scattering data
from which it was reconstructed. Basic results are stated in
Theorems 1.1 -- 1.4 below.

Consider the equation
\begin{equation}
  (I+H_x)\Gamma_x:=\Gamma_x(t,s)
  +\int^x_0H(t-u) \Gamma_x(u,s)du
  =H(t-s), \quad 0\leq t,s\leq x.
  \tag{1.6} \end{equation}
Equation (1.6) shows that $\Gamma_x=(I+H_x)^{-1}H=I-(I+H_x)^{-1}$, so
\begin{equation}
 (I+H_x)^{-1}=I-\Gamma_x \tag{1.6'}
\end{equation}
in operator form, and
\begin{equation}
H=(I-\Gamma_x)^{-1}-I.\tag{1.6"}
\end{equation} 
Let us assume that $H(t)$ is a real-valued even function
  $$ H(-t)=H(t),\quad H(t)\in L^1(\R)\cap L^2(\R), $$
\begin{equation}
  1+\tildeH(k)>0,\quad
  \tildeH(k):=\int^\infty_{-\infty} H(t)e^{ikt} dt
    =2\int^\infty_0\cos(kt)H(t)dt.
  \tag{1.7} \end{equation}
Then (1.6) is uniquely solvable for any $x>0$,
and there exists a limit
\begin{equation}
  \Gamma(t,s)=\lim_{x\to\infty} \Gamma_x(t,s):=\Gamma_{\infty}(t,s),
\quad t,s\geq 0,
  \tag{1.8} \end{equation}
where $\Gamma(t,s)$ solves the equation
\begin{equation}
  \Gamma(t,s)+\int^\infty_0 H(t-u) \Gamma(u,s)du=H(t-s),\quad
  0\leq t, s<\infty.
  \tag{1.9} \end{equation}
Given $H(t)$, one solves (1.6), finds $\Gamma_{2x}(s,0)$, then defines
\begin{equation}
  \psi(x,k):=\frac{E(x,k)-E(x,-k)}{2i},
  \tag{1.10} \end{equation}
where
\begin{equation}
  E(x,k):= e^{ikx}
   \left[ 1-\int^{2x}_0 \Gamma_{2x}(s,0) e^{-iks}ds \right].
  \tag{1.11} \end{equation}
Formula (1.11) gives a one-to-one correspondence between
$E(x,k)$ and $\Gamma_{2x}(s,0)$.
\begin{remark}
In [1]  $\Gamma_{2x}(0,s)$ is used in place
of $\Gamma_{2x}(s,0)$ in the definition of $E(x,k)$.
By formula (2.21) (see section 2 below) one has
$\Gamma_x(0,x)=\Gamma_x(x,0)$,
but $\Gamma_x(0,s)\neq \Gamma_x(s,0)$ in general.
\end{remark}
Note that
\begin{equation}
  E(x,\pm k)=e^{\pm ikx}f(\pm k)+o(1), \quad x\to+\infty,
  \tag{1.12} \end{equation}
where
\begin{equation}
  f(k):=1-\int^\infty_0 \Gamma(s) e^{iks}ds,
  \tag{1.13} \end{equation}
\begin{equation}
  \Gamma(s):=\lim_{x\to +\infty} \Gamma_x(s,0):=\Gamma_{\infty}(s,0),
  \tag{1.14} \end{equation}
and
\begin{equation}
  \psi(x,k)=\frac{e^{ikx} f(-k)-e^{-ikx}f(k)}{2i}
   +o(1),\quad x\to+\infty.
   \tag{1.15} \end{equation}
Note that
$$\psi(x,k)=|f(k)|\sin (kx+\delta(k)) + o(1), \quad x\to +\infty,
$$
where 
$$f(k)=|f(k)|e^{-i\delta(k)}, \quad \delta(k)=-\delta(-k),\quad k\in \R.$$
The function $\delta(k)$ is called the phase shift. One has $S(k)=
e^{2i\delta(k)}$.

We have changed the notations from \cite{K1} in order to
show the physical meaning of the quantity (1.12):
$f(k)$ is the Jost function of the scattering theory.
The function $\psi(x,k)/f(k)$ is the solution
to the scattering problem (see equation (2.39) below).
Krein \cite{K1} calls
\begin{equation}
  S(k):=\frac{f(-k)}{f(k)}
  \tag{1.16} \end{equation}
the $S$-function, and we will show that (1.16) is the $S$-matrix
used in physics. 

Assuming no bound states, one can solve the inverse scattering problem
(ISP):
$$ \hbox{\it given $S(k)\,\,\forall  k>0$, find $q(x)$}. $$
A solution of the ISP based on the results of \cite{K1} consists
of four steps:
\begin{description}
\item{1)} Given $S(k)$, find $f(k)$ by solving the 
Riemann problem (2.37).
\item{2)} Given $f(k)$, calculate $H(t)$ using the formula
\begin{equation}
  1+\tildeH=1+\int^\infty_{-\infty} H(t)e^{ikt}dt=\frac{1}{|f(k)|^2}.
  \tag{1.17} \end{equation}
\item{3)} Given $H(t)$, solve (1.6) and find
  $\Gamma_x(t,s)$ and then $\Gamma_{2x}(2x,0)$, $0\leq
x<\infty$.
\item{4)} Define
\begin{equation}
  A(x)=2\Gamma_{2x}(2x,0),
  \tag{1.18} \end{equation}
where
\begin{equation}
  A(0)=2H(0),
  \tag{$1.18'$}\end{equation}
and calculate the potential
\begin{equation}
  q(x)=A^2(x)+A^\prime(x), \quad A(0)=2H(0).
  \tag{1.19} \end{equation}

\end{description}
One can also calculate $q(x)$ by the formula:
$$q(x)=2\frac d {dx}[\Gamma_{2x}(2x,0)-\Gamma_{2x}(0,0)].$$
Indeed, $2\Gamma_{2x}(2x,0)=A(x),$ see (1.18),
$\frac d {dx}\Gamma_{2x}(0,0)=
-\Gamma_{2x}(2x,0)\Gamma_{2x}(0,2x),$ see (2.22), and
$\Gamma_{2x}(2x,0)=\Gamma_{2x}(0,2x),$ see (2.21).

There is an alternative (known) way, based
on the Wiener-Levy theorem, to do step 1): 

Given $S(k)$, find $\delta(k),$ the phase shift, 
then calculate the function 
$$g(t):= -\frac{2}{\pi}\int_0^{\infty}\delta(k)\sin (kt)dk,$$
and finally calculate 
$$f(k)=\exp \left(\int_0^{\infty}g(t)e^{ikt}dk\right).$$

The potential $q\in L_{1,1}$ 
generates the $S$-matrix $S(k)$ with which we started provided that
the following conditions (1.20) -(1.22) hold:
\begin{equation}
  S(k)=\overline{S(-k)}=S^{-1}(k), \quad k\in\R,
  \tag{1.20} \end{equation}
the overbar stands for complex conjugation,

and

\begin{equation}
  \ ind_\R S(k)=0,
  \tag{1.21} \end{equation}
\begin{equation}
  ||F(x)||_{L^\infty(\R_+)} +||F(x)||_{L^1(\R_+)}
   +||xF^\prime(x)||_{L^1(\R_+)}<\infty,
  \tag{1.22} \end{equation}
where
\begin{equation}
  F(x):=\frac{1}{2\pi}\int^\infty_{-\infty} [1-S(k)] e^{ikx} dk.
  \tag{1.23} \end{equation}
By the index (1.21) one means the increment of the
argument of $S(k)$ 
( when $k$ runs from $-\infty$ to $+\infty$
along the real axis) divided by $2\pi$.
The function (1.10) satisfies the equation
\begin{equation}
  \psi^{\prime\prime}+k^2\psi-q(x)\psi=0,\quad x\in\R_+.
  \tag{1.24} \end{equation}
Recall that {\it we have assumed that there are no bound states.}

In section 2 the above method is justified and the following
theorems are proved:

\begin{theorem}  
 If (1.20)-(1.22) hold, then $q(x)$ defined by (1.19) is the unique
 solution to ISP and this $q(x)$ has $S(k)$ as the scattering matrix.
\end{theorem}

\begin{theorem} 
 The function $f(k)$, defined by (1.13), is the Jost function corresponding
 to potential (1.19).
\end{theorem}

\begin{theorem} 
 Condition (1.7) implies that equation (1.6) is
 solvable for all $x\geq 0$ and its solution is unique.
\end{theorem}

\begin{theorem} 
If condition (1.7) holds, then relation
 (1.14) holds and $\Gamma(s):=\Gamma_{\infty}(s,0)$ 
is the unique solution to the
equation
\begin{equation}
  \Gamma(s)+\int^\infty_0 H(s-u)\Gamma(u)du=H(s), \quad s\geq 0.
  \tag{1.25} \end{equation}
\end{theorem}

The diagram explaining the inversion method for solving ISP, based
on Krein's results from \cite{K1}, can be shown now:
\begin{equation}
  S(k)\operatorname*\Rightarrow^{(2.38)}_{s_1}
  f(k)\operatorname*\Rightarrow^{(1.17)}_{s_2}
  H(t)\operatorname*\Rightarrow^{(1.6)}_{s_3}
  \Gamma_x(t,s)\operatorname*\Rightarrow^{\hbox{(trivial)}}_{s_4}
  \Gamma_{2x}(2x,0)\operatorname*\Rightarrow^{(1.18)}_{s_5}
  A(x)\operatorname*\Rightarrow^{(1.19)}_{s_6} q(x).
  \tag{1.26} \end{equation}
In this diagram $s_m$ denotes step number $m$. Steps $s_2$, $s_4$,
$s_5$ and $s_6$ are trivial. Step $s_1$ is almost trivial: it
requires solving a Riemann problem with index zero and can be
done analytically, in closed form. Step $s_3$ is the
basic (non-trivial) step which requires solving a family of
Fredholm-type linear integral equations (1.6).
These equations are uniquely solvable if assumption (1.7) holds, or if
assumptions (1.20)-(1.22) hold.

We analyze in section 2 the invertibility of 
the steps in diagram (1.26).
Note also that, if one assumes (1.20)-(1.22), diagram (1.26) can be
used for solving the inverse problems of finding $q(x)$ from
the following data:

\begin{description}
\item[a)] from $f(k)$, $\forall k>0$,
\item[b)] from $|f(k)|^2$, $\forall k>0$, 
or
\item[c)] from the spectral function $d\rho(\lambda)$.
\end{description}

Indeed, if (1.20)-(1.22) hold, then a) and b) are 
contained in diagram
(1.26), and c) follows from the known  formula (e.g., \cite{R1},
p.256)
\begin{equation}
  d\rho(\lambda)=\left\{ \begin{array}{rl}
  \frac{\sqrt{\lambda}}{\pi} \frac{ds} { |f(\sqrt{\lambda} ) |^2 },
    & \quad \lambda>0 \\
  0, & \quad \lambda<0.
  \end{array}\right. 
  \tag{1.27} \end{equation}
Let $\lambda=k^2$.
Then (still assuming (1.21)) one has:
\begin{equation}
  d\rho=\frac{2k^2}{\pi}
  \frac{1}{|f(k)|^2} dk, \quad k>0.
  \tag{1.28} \end{equation}

Note that the general case of the inverse scattering problem on the
half-axis, when
  $$ \ind_\R S(k):=\nu\not= 0, $$
can be reduced to the case $\nu=0$ by the procedure
described in section 4 provided that $S(k)$ is the $S-$matrix
corresponding to a potential $q\in L_{1,1}(\R_+)$.
Necessary and sufficient conditions for such an $S(k)$
are conditions (1.20)-(1.22) (see [4]).

Section 3 contains a discussion of the numerical 
aspects of the inversion
procedure based on Krein's method.
There are advantages in using this procedure (as compared with the
Gelfand-Levitan procedure): integral equation (1.6),
solving of which constitutes the basic step in the Krein inversion
method, is a Fredholm convolution-type equation.
Solving such an equation numerically leads to inversion of
Toeplitz matrices, which can be done efficiently and with much
less computer time than solving the Gelfand-Levitan equation (5.3).
Combining Krein's and Marchenko's inversion methods yields the most
efficient way to solve inverse scattering problems.

Indeed, for small $x$ equation (1.6) can be solved by iterations since
the norm of the integral operator in (1.6) is less than 1 for
sufficiently small $x$, say $0<x<x_0$.
Thus $q(x)$ can be calculated for
$0\leq x\leq \frac{x_0}{2}$ by diagram (1.26).

For $x>0$ one can solve by iterations Marchenko's equation for the
kernel $A(x,y)$:
\begin{equation}
  A(x,y) +F(x+y) +\int^\infty_x A(x,s) F(s+y)ds
  =0, \quad 0\leq x\leq y<\infty,
  \tag{1.29} \end{equation}
where, if (1.21) holds, the known function $F(x)$ is defined
by the formula:
\begin{equation}
  F(x):=\frac{1}{2\pi} \int^\infty_{-\infty} [1-S(k)] e^{ikx}dk.
  \tag{1.30} \end{equation}
Indeed, for $x>0$ the norm of the operator in (1.29) is less
than 1 (\cite{M}) and it tends to $0$ as $x\to+\infty$.

Finally let us discuss the following question: in the justification of
both the Gelfand-Levitan and Marchenko methods, the eigenfunction
expansion theorem and the Parseval relation play the fundamental
role. In contrast, the Krein method apparently does not use the
eigenfunction expansion theorem and the Parseval relation.
However, implicitly, this method is also based on such relations.
Namely, assumption (1.7) implies that the function $S(k)$, that is,
the $S$-matrix corresponding to the potential (1.19),
has index $0$. If, in addition, this potential is in
$L_{1,1}(\R_+)$, then conditions (1.20) and (1.22) are satisfied
as well, and the eigenfunction expansion theorem and 
Parseval's equality
hold. Necessary and sufficient conditions, imposed directly on the
function $H(t)$, which guarantee that conditions (1.20)-(1.22) hold,
are not known.
However, from the results of section 2 it follows that conditions
(1.20)-(1.22) hold if and only if $H(t)$ is such that the diagram
(1.26) leads to a $q(x)\in L_{1,1}(\R_+)$.
Alternatively, conditions (1.20)-(1.22) hold (and consequently,
$q(x)\in L_{1,1}(\R_+)$) if and only if condition (1.7) holds and
the function $f(k)$, which is uniquely defined as the solution
to the Riemann problem
\begin{equation}
  \Phi_+(k)=[1+\tildeH(k)]^{-1}\Phi_-(k),\quad k\in\R,
  \tag{1.31} \end{equation}
by the formula
\begin{equation}
  f(k)=\Phi_+(k),
  \tag{1.32} \end{equation}
generates the $S$-matrix $S(k)$ (by formula (1.16)), and this
$S(k)$ satisfies conditions (1.20)-(1.22).
Although the above conditions are 
verifiable in principle, they are not
satisfactory because they are implicit, they 
are not formulated in terms of
structural properties of the function $H(t)$ (such as smoothness,
rate of decay, etc.).

In section 2 Theorems 1.1 -- 1.4 are proved.
In section 3 numerical aspects of the inversion method based on
Krein's results are discussed.
In section 4 the ISP with bound states is discussed.
In section 5 a relation between Krein's and Gelfand-Levitan's
methods
is explained.

\section{Proofs}

\begin{proof}[Proof of Theorem 1.3]
If $v\in L^2(0,x)$, then
\begin{equation}
 (v+H_xv,v)=\frac{1}{2\pi}
[(\tildev,\tildev)_{L^2(\R)}+(\tildeH\tildev,\tildev)_{L^2(\R)}]
 \tag{2.1} \end{equation}
where the Parseval equality was used,
\begin{equation}
 \begin{align}
  \tildev:=&\int^x_0 v(s)e^{iks}ds,  \notag\\
  (v,v)=& \int^x_0 |v|^2ds=(\tildev,\tildev)_{L^2(\R)}.
 \tag{2.2}\end{align} \end{equation}
Thus $I+H_x$ is a positive definite selfadjoint operator in the Hilbert
space $L^2(0,x)$ if (1.7) holds. Note that, since $H(t)\in L^1(\R)$,
one has $\tildeH(k)\to 0$ as $|k|\to\infty$, so (1.7) implies
\begin{equation}
  1+\tildeH(k)\geq c>0.
  \tag{2.3} \end{equation}
A positive definite selfadjoint operator in a Hilbert space is boundedly
invertible. Theorem 1.3 is proved.
\end{proof}

Note that our argument shows that
\begin{equation}
  ||(I+H_x)^{-1}||_{L^2(\R)}\leq c^{-1}.
  \tag{2.4} \end{equation}
Before we prove Theorem 1.4, let us prove a simple  lemma.
For  results of this type, see \cite{K2}.

\begin{lemma}
The operator
\begin{equation}
  H\varphi:=\int^\infty_0 H(t-u)\varphi(u) du
  \tag{2.5} \end{equation}
is a bounded operator in $L^p(\R_+)$, $p=1,2,\infty$.

For  $\Gamma_x(u,s)\in L^1(\R_+)$ one has
\begin{equation}
||\int^\infty_x du
    H(t-u) \Gamma_x(u,s)||_{L^{2}(0,x)}\leq c_1(\int^\infty_x du
    |\Gamma_x(u,s)|)^2.
\tag{2.6}\end{equation}
\end{lemma}

\begin{proof}
Let $||\varphi||_p:=||\varphi||_{L^p(\R_+)}$.
One has
\begin{equation}
  ||H\varphi||_1\leq \sup_{u\in\R_+} \int^\infty_0 dt
    |H(t-u)| \int^\infty_0 |\varphi(u)| du
    \leq \int^\infty_{-\infty} |H(s)|ds ||\varphi||_1
    =2||H||_1\, ||\varphi||_1,
  \tag{2.7} \end{equation}
where we have used the assumption $H(t)=H(-t)$.
Similarly,
\begin{equation}
  ||H\varphi||_\infty\leq 2||H||_1\, ||\varphi||_\infty.
  \tag{2.8} \end{equation}
Finally, using Parseval's equality, one gets:
\begin{equation}
  ||H\varphi||^2_2 =2\pi||\tildeH\tilde\varphi_+||^2_{L^2(\R)}
  \leq  2\pi \sup_{k\in\R} |\tildeH(k)|^2 ||\varphi||^2_2,
  \tag{2.9} \end{equation}
where
\begin{equation}
  \varphi_+(x):=\left\{ 
     \begin{array}{rl}\varphi(x), &\ x\geq 0,\\0,& \ x<0.\end{array}\right.
  \tag{2.10}\end{equation}
Since $|\tildeH(k)|\leq 2||H||_1$ one gets from (2.9) the estimate:
\begin{equation}
  ||H\varphi||_2 \leq  2 \sqrt{2\pi}||H||_1\ ||\varphi||_2.
  \tag{2.11} \end{equation}
To prove (2.6), one notes that 
$$
\int_0^xdt|\int^\infty_x du
    H(t-u) \Gamma_x(u,s)|^2\leq 
\sup_{u,v\geq x}\int_0^xdt|H(t-u) H(t-v)|
(\int_x^{\infty}|\Gamma_x(u,s)|du)^2
$$
$$
\leq c_1(\int^\infty_x du
    |\Gamma_x(u,s)|)^2.
$$
Estimate (2.6) is obtained.
Lemma 2.1 is proved.
\end{proof}

\begin{proof}[Proof of Theorem 1.4]

Define $\Gamma_x(t,s)=0$ for $t$ or $s$ greater than $x$.
Let $w:=\Gamma_x(t,s)-\Gamma(t,s)$.
Then (1.6) and (1.9) imply

\begin{equation}
  (I+H_x)w=\int^\infty_x H(t-u)\Gamma(u,s)du:=h_x(t,s).
  \tag{2.12} \end{equation}
If condition (1.7) holds, then
 equations (1.9) and 91.25) have solutions in $L^1(\R_+)$,
and, since $\sup_{t\in \R}|H(t)|<\infty$, it is clear that
 this solution belongs to $L^\infty(\R_+)$ and consequently
to $L^2(\R_+)$, because $||\varphi||_2\leq 
||\varphi||_\infty||\varphi||_1$.
The proof of Theorem 1.3 shows that such a solution is unique and
does exist. From (2.4) one gets
\begin{equation}
  \sup_{x\geq 0} ||(I+H_x)^{-1}||_{L^2(0,x)} \leq c^{-1}.
  \tag{2.13} \end{equation}
For any fixed $s>0$ one sees that $\sup_{x\geq y}||h_x(t,s)||\to
0$ as $y\to \infty$, where the norm here stands for any of the three
norms $L^p(0,x), p=1,2,\infty$.
Therefore (2.12) and (2.11) imply
\begin{equation}
  \begin{align}
  ||w||^2_{L^2(0,x)} &\leq c^{-2} ||h_x||_{0,x}\notag\\ 
   &\leq c^{-2}
  \left\| \int^\infty_x H(t-u) \Gamma(u,s)dy\right\|_{L^1(0,x)}
  \left\| \int^\infty_x H(t-u) \Gamma(u,s)dy\right\|_{L^\infty(0,x)}
  \notag\\
  &\leq \hbox{\ const\ }
  \left\| \Gamma(u,s)\right\|^2_{L^1(x,\infty)}\to 0
  \hbox{\ as\ }x\to\infty,
  \tag{2.14}  \end{align}
  \end{equation}
since $\Gamma(u,s)\in L^1(\R_+)$ for any fixed $s>0$ and 
$H(t)\in L^1(\R)$.

Also
\begin{equation}
  \begin{align}
  \|w(t,s)\|^2_{L^\infty(0,x)} \leq 2(||h_x||^2_{L^\infty(0,x)}
+||H_xw||^2_{L^\infty(0,x)})\leq \notag \\
 c_1 ||\Gamma (u,s)||^2_{L^1(x,\infty)}+
c_2 \sup_{t\in \R}||H(t-u)||^2_{L^2(0,x)} ||w||^2_{L^2(0,x)},
  \tag{2.15} \end{align} \end{equation}
where $c_j>0$ are some constants. 
Finally, by (2.6), one has;
\begin{equation}
  \begin{align}
  \| w(t,s)\|^2_{L^2(0,x)}
 &  \leq c_3 (\int^\infty_x  |\Gamma(u,s)|du)^2\to 0
  \hbox{\ as\ } x\to+\infty.
  \tag{2.16} \end{align}\end{equation}
From (2.15) and (2.16) relation (1.14) follows.
Theorem 1.4 is proved.
\end{proof}

Let us now prove Theorem 1.2.
We need several lemmas.

\begin{lemma}
The function (1.11) satisfies the equations
\begin{equation}
  E^\prime=ikE - A(x)E_-, \quad E(0,k)=1, \quad E_-:=E(x,-k),
  \tag{2.17} \end{equation}
\begin{equation}
  E^\prime_-=-ikE_- -A(x)E, \quad E_-(0,k)=1,
  \tag{2.18} \end{equation}
where $E^\prime =\frac{dE}{dx}$,
and $A(x)$ is defined in (1.18).
\end{lemma}

\begin{proof}
Differentiate (1.11) and get
\begin{equation}
  E^\prime =ikE-e^{ikx} \left(
    2\Gamma_{2x} (2x,0)e^{-ik2x}+2\int^{2x}_0
    \frac{\partial \Gamma_{2x}(s,0)}{\partial (2x)} e^{-iks} ds
    \right) \tag{2.19} \end{equation}
We will check below that
\begin{equation}
  \frac{\partial\
  \Gamma_{x}(t,s)}{\partial x} =-\Gamma_x(t,x)\Gamma_x(x,s)
  \tag{2.20} \end{equation}
and
\begin{equation}
  \Gamma_x(t,s)=\Gamma_x(x-t, x-s).
  \tag{2.21} \end{equation}
Thus, by (2.20),
\begin{equation}
  \frac{\partial \Gamma_{2x}(s,0)}{\partial(2x)} =-\Gamma_{2x}
  (s,2x)\Gamma_{2x}(2x,0).
  \tag{2.22}\end{equation}
Therefore (2.19) can be written as
\begin{equation}
  E^\prime =ikE-e^{-ikx} A(x)+A(x)e^{ikx}
  \int^{2x}_0 \Gamma_{2x}(s,2x) e^{-iks}ds.
  \tag{2.23} \end{equation}

By (2.21) one gets
\begin{equation}
  \Gamma_{2x}(s,2x)=\Gamma_{2x}(2x-s,0).
  \tag{2.24} \end{equation}
Thus
\begin{equation}
  \begin{align}
  e^{ikx} \int^{2x}_0 \Gamma_{2x} (s,2x) e^{-iks} ds
 & =\int^{2x}_0\Gamma_{2x}(2x-s,0)e^{ik(x-s)} ds
 \notag\\
 & =e^{-ikx} \int^{2x}_0 \Gamma_{2x}(y,0) e^{iky} dy.
  \tag{2.25} \end{align}
  \end{equation}
From (2.23) and (2.25) one gets (2.17).

Equation (2.18) can be obtained from (2.17) by changing $k$ to $-k$.
Lemma 2.2 is proved if formulas (2.20)-(2.21) are checked.

To check (2.21), use $H(-t)=H(t)$ and compare the equation for
$\Gamma_x(x-t,x-s):=\varphi$,
\begin{equation}
  \Gamma_x(x-t,x-s)+\int^x_0H(x-t-u)\Gamma_x(u,x-s)du
  =H(x-t-x+s)=H(t-s),
  \tag{2.26} \end{equation}
with equation (1.6). Let $u=x-y$. Then (2.26) can be written as
\begin{equation}
  \varphi+\int^x_0 H(t-y)\varphi\,dy =H(t-s),
  \tag{2.27} \end{equation}
which is equation (1.6) for $\varphi$. Since (1.6) has at most one
solution, as we have proved above, formula (2.21) is proved.

To prove (2.20), differentiate (1.6) with respect to $x$ and get:
\begin{equation}
  \Gamma^\prime_x(t,s)+\int^x_0 H(t-u) \Gamma^\prime_x(u,s)du
  =-H(t-x)\Gamma_x(x,s)
  \tag{2.28} \end{equation}
Set $s=x$ in (1.6), multiply (1.6) by $-\Gamma_x(x,s)$,
compare with (2.28) and use again the uniqueness of the solution
to (1.6). This yields (2.20).

Lemma 2.2 is proved.
\end{proof}

\begin{lemma}
Equation (1.24) holds.
\end{lemma}

\begin{proof}
From (1.10) and (2.17)-(2.18) one gets
\begin{equation}
  \psi^{\prime\prime}=\frac{E^{\prime\prime}-E^{\prime\prime}_-}{2i}
 = \frac{(ikE-A(x)E_-)^\prime -(-ikE_- -A(x)E)^\prime}{2i}.
  \tag{2.29} \end{equation}
Using (2.17)-(2.18) again one gets
\begin{equation}
  \psi^{\prime\prime}=-k^2\psi+q(x)\psi,
  \quad q(x):=A^2(x)+A^\prime(x).
  \tag{2.30} \end{equation}

Lemma 2.3 is proved.
\end{proof}

\begin{proof}[Proof of Theorem 1.2]
The function $\psi$ defined in (1.10) solves equation (1.24) and satisfies
the conditions
\begin{equation}
  \psi(0,k)=0,\quad \psi^\prime(0,k)=k.
  \tag{2.31}\end{equation}
The first condition is obvious (in \cite{K1} there is a misprint: it is written
that $\psi(0,k)=1$),
and the second condition follows from (1.10) and (2.15):
\begin{equation}
  \psi^\prime(0,k)=
  \frac{E^\prime(0,k)-E^\prime_-(0,k)}{2i}
  =\frac{ikE-AE_--(ikE_- -AE)}{2i} \bigg|_{x=0}
  =\frac{2ik}{2i}=k.
  \notag\end{equation}
Let $f(x,k)$ be the Jost solution to (1.24) which is uniquely
defined by the asymptotics
\begin{equation}
  f(x,k)=e^{ikx}+o(1), \quad x\to +\infty.
  \tag{2.32} \end{equation}
Since $f(x,k)$ and $f(x,-k)$ are lineraly independent, one has
\begin{equation}
  \psi=c_1f(x,k)+c_2f(x,-k),
  \tag{2.33}\end{equation}
where $c_1$, $c_2$ are some constants independent of $x$
but depending on $k$.

From (2.31) and (2.33) one gets
\begin{equation}
  c_1=f(-k), \quad c_2=-f(k); \quad f(k):=f(0,k).
  \tag{2.34} \end{equation}
Indeed, the choice of $c_1$ and $c_2$ guarantees that
the first condition (2.31) is obviously satisfied, while
the second follows from the Wronskian formula:
\begin{equation}
  f^\prime(0,k)f(-k)-f(k)f^\prime(0,k)=2ik.
  \tag{2.35} \end{equation}
From (2.32), (2.33) and (2.34) one gets:
\begin{equation}
  \psi(x,k)=e^{ikx} f(-k)+e^{-ikx} f(k)+o(1),
  \quad x\to +\infty.
  \tag{2.36} \end{equation}
Comparing (2.36) with (1.15) yields the conclusion of Theorem 1.2.
\end{proof}

\section*{Invertibility of the steps of the inversion
procedure and proof of Theorem 1.1}
Let us start with a discussion of the inversion
steps 1) -- 4) described in the introduction.

Then we discuss the uniqueness of the solution to ISP
and the consistency of the inversion method, that is, 
the fact that $q(x)$, reconstructed from $S(k)$ by steps 1) -- 4),
generates the original $S(k)$.

Let us go through steps 1) -- 4) of the reconstruction method and
prove their invertibility.
The consistency of the inversion method follows from the invertibility
of the steps of the inversion method.

\vskip.15in
\underbar{Step 1.}  \quad $S\Rightarrow f(k)$.

Assume $S(k)$ satisfying (1.20)-(1.22) is given. 
Then solve the Riemann
problem
\begin{equation}
  f(k)=S(-k)f(-k), \qquad k\in\R.
  \tag{2.37} \end{equation}
Since $\ind_\R S(k)=0$, one has $\ind_\R S(-k)=0$.
Therefore the problem (2.37) of finding an analytic function $f_+(k)$
in $\C_+:=\{k:\Im k>0\}$,  $f(k):=f_+(k)$ in $\C_+$, and
and analytic function $f_-(k):=f(-k)$ in
$\C_-:=\{k:\Im k<0\}$ from equation (2.37) 
can be solved in closed form.
Namely, define
\begin{equation}
  f(k)= \exp
  \left\{ \frac{1}{2\pi i} \int^\infty_{-\infty}
     \frac{\ln S(-y)dy}{y-k}\right\}, \quad \Im k>0.
  \tag{2.38} \end{equation}
Then $f(k)$ solves (2.37), $f_+(k)=f(k)$, $f_-(k)=f(-k)$.
Indeed,
\begin{equation}
  \ln\, f_+(k)-\ln\,f_-(k)=\ln\, S(k),\quad k\in\R
  \tag{2.39} \end{equation}
by the known jump formula for the Cauchy integral. Integral (2.38)
converges absolutely at infinity and $\ln\,S(-y)$
is differentiable with respect to $y$, so the Cauchy integral in (2.38) is
well defined.

To justify the above claims, one uses the known properties of the
Jost function
\begin{equation}
  f(k)=1+\int^\infty_0 A(0,y) e^{iky}dy
  := 1+\int^\infty_0 A(y) e^{iky}dy,
  \tag{2.40} \end{equation}
where (see \cite{M})
\begin{equation}
  |A(y)| \leq c \int^\infty_{\frac{y}{2}} |q(t)|dt,
  \tag{2.41} \end{equation}
\begin{equation}
  \left| \frac{\partial A(y)}{\partial y} +\frac{1}{4}
    q\left(\frac{y}{2}\right) \right|
  \leq c \int^\infty_{\frac{y}{2}} |q(t)| dt,
  \tag{2.42} \end{equation}
 $c>0$ stand for various constants and $A(y)$ is a
real-valued function. Thus
\begin{equation}
  f(k)=1-\frac{A(0)}{ik} -\frac{1}{ik}
  \int^\infty_0 A^\prime(t) e^{ikt}dt,
  \tag{2.43} \end{equation}
\begin{equation}
 S(-k)
 =\frac{f(k)}{f(-k)}
  =\frac{1-\frac{A(0)}{ik} -\frac{1}{ik} \widetilde{A^\prime}(k)}
   {1+\frac{A(0)}{ik} + \frac{1}{ik} \widetilde{A^\prime}(-k)}
   =1+0\left( \frac{1}{k}\right).
   \tag{2.44} \end{equation}
Therefore
\begin{equation}
  \ln S(-k)=0\left(\frac{1}{k}\right)
  \quad\hbox{as}\quad |k|\to\infty, \quad k\in\R.
  \tag{2.45}\end{equation}
Also
\begin{equation}
  \dotf(k)=1+i \int^\infty_0 A(y)y e^{iky}dy,
  \quad \dotf:=\frac{\partial f}{\partial k}.
  \tag{2.46} \end{equation}
Estimate (2.41) implies
\begin{equation}
  \int^\infty_0 y|A(y)| dy\leq 2\int^\infty_0 t|q(t)|dt<\infty,
  \tag{2.47} \end{equation}
so that $f^\prime(k)$ is bounded for all $k\in\R$,
as claimed. Note that
\begin{equation}
  f(-k)=\overline{f(k)}, \quad k\in\R.
  \tag{2.48} \end{equation}
The converse step
$$ f(k)\Rightarrow S(k) $$
is trivial:
  $$ S(k)=\frac{f(-k)}{f(k)}. $$
If $\ind_\R S=0$ then $f(k)$ is analytic in $\C_+$,
$f(k)\not= 0$ in $\C_+$, $f(k)=1+O\left(\frac{1}{k}\right)$
as $|k|\to\infty$, $k\in\C_+$, and (2.48) holds.

\vskip.15in
\underbar{Step 2.} \quad $f(k)\Rightarrow H(t)$.

This step is done by formula (1.17):
\begin{equation}
  H(t)=\frac{1}{2\pi} \int^\infty_{-\infty} e^{-ikt}
  \left( \frac{1}{|f(k)|^2} -1\right) dk.
  \tag{2.49} \end{equation}
The integral in (2.49) converges in $L^2(\R)$.
Indeed, it follows from (2.43) that
\begin{equation}
 |f(k)|^2-1 =-\frac{2}{k} \int^\infty_0 A^\prime(t) \sin(kt)dt
 +O\left( \frac{1}{|k|^2}\right),
 \quad |k|\to\infty,\quad k\in\R.
 \tag{2.50} \end{equation}
The function
\begin{equation}
  w(k):=\frac{1}{k} \int^\infty_0 A^\prime(t) \sin(kt)dt
  \tag{2.51} \end{equation}
is continuous because $t A^\prime(t)\in L^1(\R_+)$
by (2.42) and $w\in L^2(R)$ since $w=o\left(\frac{1}{|k|}\right)$
as $|k|\to\infty$, $k\in\R$.

The converse step
\begin{equation}
  H(t)\Rightarrow f(k)
  \tag{2.52} \end{equation}
is also done by formula (1.17): Fourier inversion
gives $|f(k)|^2=f(k)f(-k),$ and factorization yields the unique $f(k)$,
since $f(k)$ does not vanish in $\C_+$ and tends to $1$ at infinity.

\vskip.15in
\underbar{Step 3.} \quad $H\Rightarrow \Gamma_x(s,0)\Rightarrow
 \Gamma_{2x}(2x,0).$

This step is done by solving equation (1.6). By Theorem 1.3
equation (1.6) is uniquely solvable since condition (1.7) is assumed.
Formula (1.17) holds and the known properties of the Jost function
(1.4) are used:
$f(k)\to 1$ as $k\to\pm\infty$, $f(k)\not=0$ for $k\not= 0$, $k\in\R$,
$f(0)\not= 0$ since $\ind_\R S(k)=0$.

The converse step $\Gamma_x(s,0)\Rightarrow H(t)$ is done
by formula (1.6"). The converse step
\begin{equation}
  \Gamma_{2x}(2x,0)\Rightarrow \Gamma_x(s,0)
  \tag{2.53} \end{equation}
constitutes the essence of the inversion method.

This step is done as follows:
\begin{equation}
  \Gamma_{2x}(2x,0)\operatorname*\Rightarrow^{(1.18)}
  A(x) \operatorname*\Rightarrow^{(2.17)-(2.18)}
  E(x,k) \operatorname*\Rightarrow^{(1.11)}
  \Gamma_x(s,0). 
  \tag{2.54} \end{equation}
Given $A(x)$, system (2.17)-(2.18) is uniquely solvable for $E(x,k)$.

Note that the step $q(x)\Rightarrow f(k)$ can be done by solving the
uniquely

solvable integral equation (see \cite{R1}):
\begin{equation}
  f(x,k)=e^{ikx} +\int^\infty_x
  \frac{\sin[k(y-x)]}{k}q(y)f(y,k)dy
  \tag{2.55} \end{equation}
with $q\in L_{1,1}(\R_+)$, and then calculating
$f(k)=f(0,k)$.

\vskip.15in
\underbar{Step 4.} \quad $A(x):=\Gamma_{2x}(2x,0)\Rightarrow q(x).$

This step is done by formula (1.19). The converse step
$$
q(x)\Rightarrow A(x)
$$
can be done by solving the Riccati problem (1.19) for $A(x)$
given $q(x)$ and the initial condition $2H(0)$. 
Given  $q(x)$, one can find $2H(0)$ as follows: 
one finds $f(x,k)$ by solving equation (2.55),
which is uniquely solvable if $q \in L_{1,1}(\R_+)$, then
one gets $f(k):=f(0,k)$, and then calculates 
$2H(0)$ using formula (2.49) with $t=0$:
$$
2H(0)= \frac{1}{\pi} \int^\infty_{-\infty} 
  \left( \frac{1}{|f(k)|^2} -1\right) dk.
$$

\begin{proof}[Proof of Theorem 1.1.]
If (1.20)-(1.22) hold, then, as has been proved in \cite{R1}
(and earlier in a different form in \cite{M}), there is a unique
$q(x)\in L_{1,1}(\R_+)$ which generates the given
$S$-matrix $S(k)$.

It is not proved in \cite{K1} that $q(x)$ defined in (1.19)
(and obtained as a final result of steps 1) -- 4))
generates the scattering matrix $S(k)$ with which we started
the inversion. 

Let us now prove this.
We have already discussed the following diagram:
\begin{equation}
  S(k)\operatorname*\Leftrightarrow^{(2.38)}_{(1.16)}
  f(k)\operatorname*\Leftrightarrow^{(1.17)}
  H(t)\operatorname*\Leftrightarrow^{(1.6)}
  \Gamma_x(s,0)\operatorname*\Rightarrow\Gamma_{2x}(2x,0)
\operatorname*\Leftrightarrow^{(1.18)}
  A(x)\operatorname*\Leftrightarrow^{(1.19)} q(x).
  \tag{2.56} \end{equation}
To close this diagram and therefore establish the basic one-to-one
correspondence
\begin{equation}
  S(k)\Leftrightarrow q(x)
  \tag{2.57} \end{equation}
one needs to prove
$$\Gamma_{2x}(2x,0)\Rightarrow\Gamma_x(s,0).$$ 

This is done by the scheme (2.54).

Note that the step
$q(x)\Rightarrow A(x)$ requires solving Riccati equation (1.19)
with the boundary condition $A(0)=2H(0)$.
Existence of the solution to this problem on all of $\R_+$ is
guaranteed by the assumptions (1.20)-(1.22). The fact
that these assumptions imply $q(x)\in L_{1,1}(\R_+)$ is proved
in \cite{M} and \cite{R1}.
Theorem 1.1 is proved.
\end{proof}

Uniqueness theorems for the inverse scattering problem are not given
in \cite{K1}, \cite{K2}. They can be found in \cite{M}-\cite{R4}.

\begin{remark}
From our analysis one gets the following result:
\end{remark}

\begin{proposition}

If $q(x)\in L_{1,1}(\R_+)$ and has no bounds states and no resonance at
zero, then Riccati equation (1.19) with the initial condition (1.18)
has the solution $A(x)$ defined for all $x\in\R_+$.
\end{proposition}

\section{Numerical aspects of the Krein inversion procedure.}
The main step in this procedure from the numerical viewpoint is to
solve equation (1.6) for all $x>0$ and all $0<s<x$, which are the
parameters in equation (1.6).

Since equation (1.6) is an equation with the convolution kernel, its
numerical solution involves inversion of a Toeplitz matrix,
which is a well developed area of numerical analysis.
Moreover, such an inversion requires much less computer 
memory and time than the inversion based on
the Gelfand-Levitan or Marchenko methods.
This is the main advantage of Krein's inversion method.

This method may become even more attractive if it is combined with
the Marchenko method. In the Marchenko method the equation to be solved
is (\cite{M}, \cite{R1}):
\begin{equation}
  A(x,y)+F(x+y)+\int^\infty_x A(x,s)F(s+y)ds=0,
  \quad y\geq x\geq 0,
  \tag{3.1} \end{equation}
where $F(x)$ is defined in (1.23) and is known if $S(k)$ is known,
the kernel $A(x,y)$ is to be found from (3.1) and
if $A(x,y)$ is found then the potential is recovered by the formula:
\begin{equation}
  q(x)=-2\frac{d A(x,x)}{dx}.
  \tag{3.2} \end{equation}
Equation (3.1) can be written in operator form:
\begin{equation}
  (I+F_x) A= -F.
  \tag{3.3} \end{equation}
The operator $F_x$ is a contraction mapping in the Banach space
$L^1(x,\infty)$ for $x>0.$
The operator $H_x$ in (1.6) is a contraction mapping in
$L^\infty(0,x)$ for $0<x<x_0$, where $x_0$ is chosen to that
\begin{equation}
  \int^{x_0}_0 |H(t-u)| du<1.
  \tag{3.4} \end{equation}
Therefore it seems reasonable from the numerical point of view to use
the following approach:

\begin{description}
\item[1.] Given $S(k)$, calculate $f(k)$ and $H(t)$ as explained
in Steps 1 and 2, and also $F(x)$ by formula (1.23).
\item[2.] Solve by iterations equation (1.6) for
$x<x<x_0$, where $x_0$ is chosen so that the iteration method
for solving (1.6) converges rapidly. Then find $q(x)$ as explained in
Step 4.
\item[3.] Solve equation (3.1) for $x>x_0$ by iterations.
Find $q(x)$ for $x>x_0$ by formula (3.2).
\end{description}

\section{Discussion of the ISP when the bound states are present.}

If the data are (1.1) then one defines
\begin{equation}
  w(k)=\prod^J_{j=1}
  \frac{k-ik_j}{k+ik_j}\hbox{\quad if\quad }\ind_\R S(x)=-2J
  \tag{4.1} \end{equation}
and
\begin{equation}
  W(k)=\frac{k}{k+i\gamma} w(k)\hbox{\quad if\quad }\ind_R S(k)=-2J-1,
  \tag{4.2} \end{equation}
where $\gamma>0$ is chosen so that $\gamma\not=k_j$,
$1\leq j\leq J$.

Then one defines
\begin{equation}
  S_1(k):=S(k) w^2(k)\hbox{\quad if\quad }\ind_\R S=-2J
  \tag{4.3} \end{equation}
or
\begin{equation}
  S_1(k):= S(k) W^2(k)\hbox{\quad if\quad }\ind_\R S=-2J-1.
  \tag{4.4} \end{equation}
Since $\ind_\R w^2(k)=2J$ and $\ind_\R W^2(k)=2J+1$, one has
\begin{equation}
  \ind_\R S_1(k)=0.
  \tag{4.5} \end{equation}
The theory of section 2 applies to $S_1(k)$ and yields $q_1(x)$.
From $q_1(x)$ one gets $q(x)$ by adding bound states $-k_j^2$ and
norming constants $s_j$ using the known procedure (e.g. see \cite{M}).

\section{Relation between Krein's and GL's methods.}

The GL (Gelfand-Levitan) method consists of the following steps
(see \cite{R1}, for example):

\vskip.15in
\underbar{Step 1.}
Given $f(k)$, the Jost function, find
\begin{equation}
  \begin{align}
  L(x,y)& :=\frac{2}{\pi} \int^\infty_0 dk\, k^2
  \left( \frac{1}{|f(k)|^2}-1\right)
  \frac{\sin kx}{k}\frac{\sin ky}{k}
  \notag\\
  &=\frac{1}{\pi} \int^\infty_0 dk\left(|f(k)|^{-2}-1\right)
  \left( \cos[k(x-y)]-\cos[k(x+y)]\right)
  \notag\\
  &:=M(x-y)-M(x+y),
  \tag{5.1} \end{align} \end{equation}
where
\begin{equation}
  M(x):=\frac{1}{\pi} \int^\infty_0dk
  \left(|f(k)|^{-2}-1\right) \cos(kx).
  \tag{5.2} \end{equation}

\vskip.15in
\underbar{Step 2.}
Solve the integral equation for $K(x,y)$:
\begin{equation}
  K(x,y)+L(x,y)+\int^x_0 K(x,s)L(s,y)ds=0,
  \quad 0\leq y\leq x.
  \tag{5.3} \end{equation}

\vskip.15in
\underbar{Step 3.}
Find
\begin{equation}
  q(x)=2\frac{d K(x,x)}{dx}.
  \tag{5.4} \end{equation}
Krein's function (see (1.17)) can be written as follows:
\begin{equation}
  H(t)=\frac{1}{2\pi} \int^\infty_{-\infty}
  \left( |f(k)|^{-2}-1 \right) e^{-ikt}dk
  =\frac{1}{\pi} \int^\infty_0 \left(|f(k)|^{-2}-1\right)
  \cos(kt)dk.
  \tag{5.5} \end{equation}

Thus the relation between the two methods is given by the 
known formula:
\begin{equation}
  M(x)=H(x).
  \tag{5.6} \end{equation}

In fact, the GL method deals with the inversion of the spectral
foundation $d\rho$ of the operator $-\frac{d^2}{dx^2}+q(x)$
defined in $L^2(\R_+)$ by the Dirichlet boundary condition
at $x=0$. However, if $\ind_\R S(k)=0$ (no bound states and no
resonance at $k=0$), then (see e.g. \cite{R1}):
\begin{equation}
  d\rho=\left\{
  \begin{array}{rl}
  \frac{2k^2dk}{\pi |f(k)|^2}, &\quad\lambda>0,\quad\lambda=k^2,\notag\\
  0, & \quad \lambda<0,
 \notag\end{array} \right. \end{equation}
so $d\rho$ in this case is uniquely defined by $f(k)$.

\end{document}